\documentclass{elsart}
\usepackage{epsf}
\usepackage{epsfig}
\usepackage{cite}

\newcommand{\beq}{\begin{equation}}
\newcommand{\eeq}{\end{equation}  }

\newcommand{\bec}{\begin{center}}
\newcommand{\eec}{\end{center}}

\def\p+p{\pi^{\pm}\mbox{p}}
\def\K+p{\mbox{K}^{+}\mbox{p}} 
\def\m+p{\mu^{+}\mbox{p}}

\newcommand{\cq}{\mathrm{c}}
\newcommand{\cqb}{\bar{\mathrm{c}}}
\newcommand{\q}{\mathrm{q}}
\newcommand{\gl}{g}

\newcommand{\bq}{\mathrm{b}}
\newcommand{\uq}{\mathrm{u}}
\newcommand{\uqb}{\bar{\mathrm{u}}}
\newcommand{\bqb}{\bar{\mathrm{b}}}
\newcommand{\Z}{\mathrm{Z}}

\makeatletter
   \newcommand\figcaption{\def\@captype{figure}\caption}
   \newcommand\tabcaption{\def\@captype{figure}\caption}
\makeatother

\begin{document}
\begin{frontmatter}

\hfill {\bf ANL-HEP-PR-00-46}

\title{\Large\bf Soft-gluon angular screening in \\  
heavy-quark fragmentation } 

\author[Argon]{S. V. Chekanov 
} 
\address[Argon]{HEP Division, Argonne National Laboratory,
9700 S.Cass Avenue, \\ 
Argonne, IL 60439
USA.\\  Email: chekanov@mail.desy.de}

\begin{abstract}
A method to measure the suppression
of soft-gluon radiation by a heavy quark (``dead cone'') is discussed. 
We analyse this QCD phenomenon in the framework of the HERA experiment
using Monte Carlo simulations. 
\end{abstract}

\bigskip 

\centerline{PACS: 12.38.Qk, 12.38.Aw}

\end{frontmatter}

\section{Introduction}
The suppression of bremsstrahlung 
off an  accelerating  massive particle 
is well known effect in classical electrodynamics \cite{tam}.
The characteristic feature of such  radiation is a 
large value of the photon  emission angle with respect to the
direction of motion of a charged particle. 
This angle is of  order  $m/E$, with $m$ and $E$
being the mass and the energy of the radiating particle.     

Similarly, this effect is expected in QCD  \cite{dro}.
The gluon radiation from heavy quarks, $\cq$ or $\bq$, is characterised  by   
the angular screening, i.e. the soft gluon emission  in the forward direction
of a  heavy quark is  reduced   
within the angle $\Theta_0 =m_q/E_q$ ($m_q$ and $E_q$ are  
the mass and the energy of the heavy quark) \cite{dok,dok1}. 
The gluon-emission probability $\mathrm{d}\sigma_{\q\to \gl\q}$ 
as a function of the angle $\Theta$ between the direction of motion of a 
soft gluon and an  emitting quark is  proportional to \cite{dok1} 
\beq
\frac{C_{\mathrm{F}}\alpha_s}{\pi}
\frac{H^2(\Theta )\mathrm{d}H^2(\Theta )}{(H^2(\Theta )+
\Theta_0^2)^2}, \qquad H(\Theta )=2\sin\Theta /2. 
\label{1}
\eeq
Thus, $\mathrm{d}\sigma_{\q\to \gl\q}$ is
suppressed for  $\Theta < \Theta_0 <<1$. This   
screening of the collinear singularity by the heavy-quark mass,
known as the ''dead cone``,  
is the most important phenomenon which determines the
shape of jets initiated by heavy quarks. For example,
the form of $\ln (1/x)$ spectra can change drastically in
the hard momenta region,  compared with jets 
initiated by light quarks \cite{dok}. A recent theoretical overview
of the dead cone effect is given in \cite{dokr,ochs}.     

Although the angular screening  has not been observed directly in QCD,
unlike to electron  radiation in QED, 
this  effect has received  an attention in Monte Carlo (MC) models
simulating main aspects of high-energy interactions.
An approximate method to include the dead cone was discussed
in \cite{mar} and has been implemented in the parton shower
of the HERWIG model \cite{HRW}.
The ARIADNE model has different untested options to reproduce the effect
in the Color Dipole Model \cite{ARD}. 
The PYTHIA/JETSET model \cite{JET} contains the dead cone by construction 
of the parton shower. Note that the effect is not exactly implemented, 
but is rather a consequence of how the shower 
kinematics is set up \cite{Sro}.

There are a few obstacles to observe the screening effect experimentally.
A large sample of  high purity $\cq$ or $\bq$ events is necessary
which then has to be used to determine the direction of the original
heavy  quarks. The dead cone can  be rather
small, thus it  is important  to have good two-track resolution. 
The decay products of  heavy mesons should be removed
from the events. The  most serious concern, however,  
is to understand how strong high-order QCD and  hadronisation 
can change hadronic angular distributions near the heavy-quark direction. 
For example, intensive color flows at hadronisation stage can
produce a smearing effect both for the reconstruction of the heavy-quark
direction as well as for the final-state hadrons originating from 
soft-gluon emissions. Resonance decay products
can  further mask the signal.

At LEP1 energies,  heavy quarks are mostly coming from 
$\Z\to \cq\cqb,\bq\bqb$ decay and carry about half of the beam energy.
Therefore,  the dead cone should exist 
at $\Theta_0\sim 2^o$ for $\cq$ quarks and 
$\Theta_0\sim 6^o$ for $\bq$ quarks \cite{mar}.
The smallness of $\Theta_0$ at LEP makes it rather
complicated to detect the angular screening  due to 
the smearing effects described above. 
Some ideas on how to find this effect at LEP have been discussed in
\cite{lepd,mar}, however, since then no progress has been made
to  determine the size of the dead cone experimentally.

HERA provides an unique opportunity to observe the soft 
gluon depletion in the charm fragmentation since 
the energy for the $\cq\cqb$ production is small.  
For typical HERA kinematics  and cuts used to 
reconstruct, say  $D^*$  mesons, the energy
of the $\cq$ ($\cqb$) quark is about 3-5 GeV. Thus the angular
screening can be observed at 
an angle of order $20^o - 30^o$,  which is by  
factor ten larger than that expected at LEP.    
As we will show in this paper, the high 
luminosity data for deep  inelastic scattering (DIS) 
and photoproduction, which is already delivered by HERA, 
allows a measurement  of this phenomenon in details.

\begin{figure}
\begin{center}
\mbox{\epsfig{file=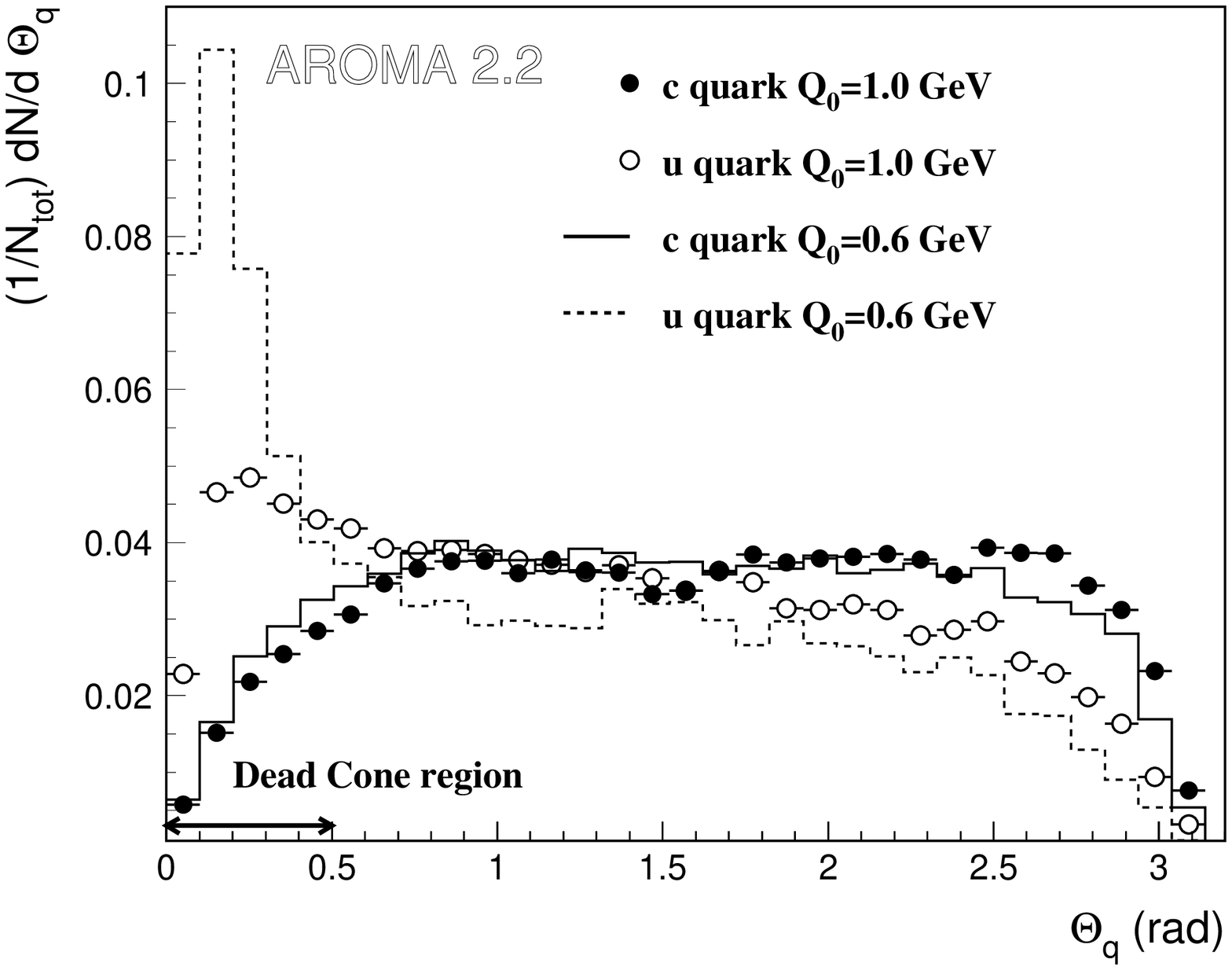,height=10.0cm}}
\caption{\it
Distributions  of the angle between  
the first parton-shower gluon emission
and $\cq$ ($\uq$)  quark (or antiquarks)
from the BGF charm production generated with AROMA for different
cut-off values $Q_0$ in the parton shower. All histograms
are normalized to unity.} 
\label{see1}
\mbox{\epsfig{file=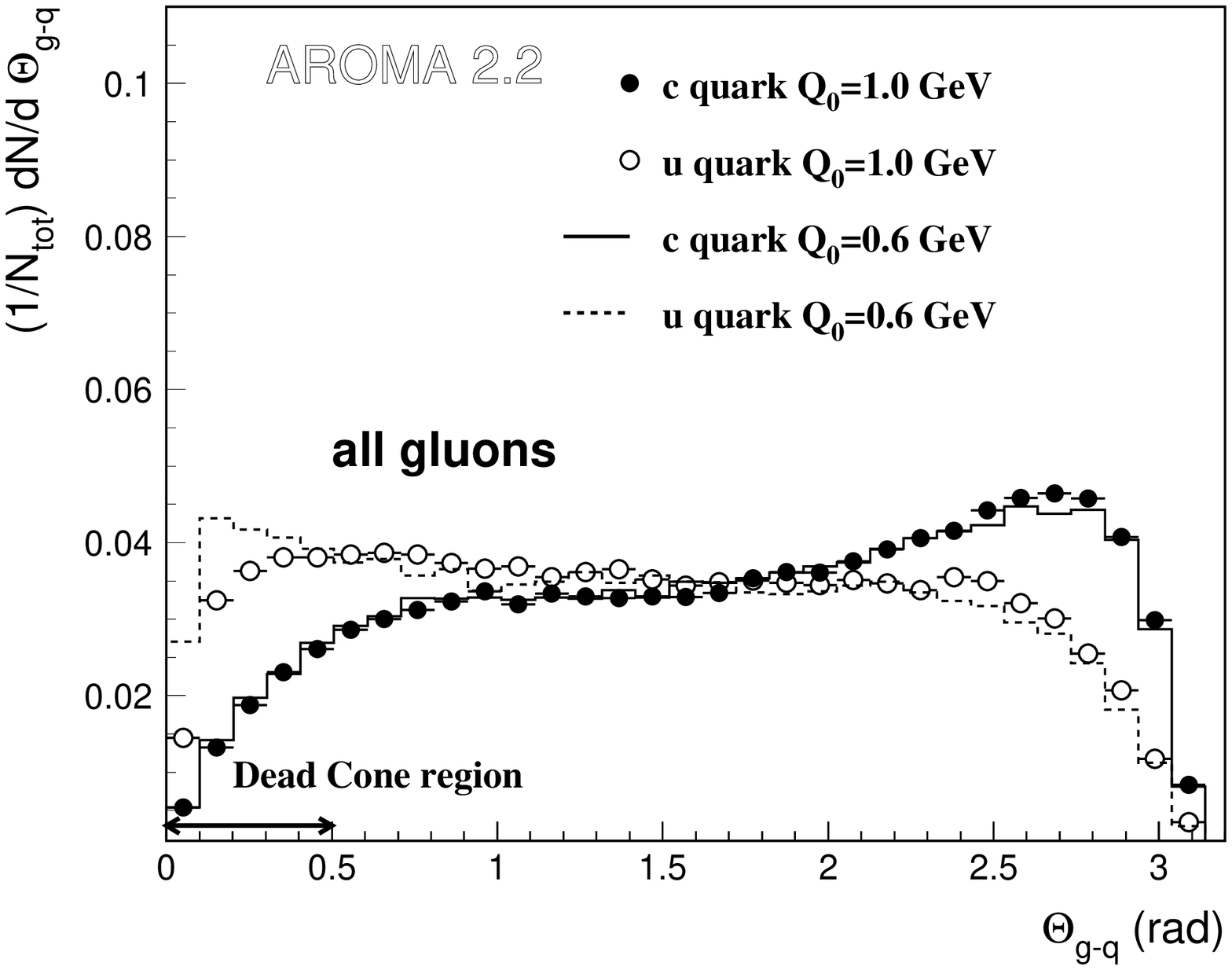,height=10.0cm}}
\caption{\it
Angular distributions of  all parton-shower gluons
with respect to the initial $\cq$ ($\uq$)  
quark (antiquarks) in the BGF
for different values of the cut-off $Q_0$.}
\label{see2}
\end{center}
\end{figure}

\section{Angular distributions}

In this paper we 
investigate  the possibility to determine the size of  the
dead cone in DIS at HERA.
For our study we  use the AROMA 2.2 Monte Carlo program \cite{ARO}
unless otherwise stated. 
This model is the most 
suitable for charm study as it contains an exact
matrix element for the heavy-quark production. 
The DIS events were generated at $Q^2>5$ GeV$^2$  with 
positron and proton beams at energies
$E_e=27.5$ GeV and $920$ GeV, respectively.
The GRV94 structure function was used.  

The AROMA generates the charm production  
only through the boson-gluon fusion (BGF) mechanism.
This model contains the initial- and  final-state QCD
radiations. The parton shower, matched to the first-order     
matrix elements on the basis of the LEPTO model \cite{LEP}, 
is simulated  with the PYTHIA/JETSET program. As mentioned above, 
the dead cone  is not exactly implemented and comes out by 
construction of the shower.   
A typical opening angle $\Theta_{\q}$ 
between gluon  $\gl$ and quark $\q$ in the splitting
$\q\to \gl\> \q$  approximately equals to \cite{JET}  
\beq
\Theta_{\q}  \simeq \frac{1}{\sqrt{z_{\q} (1-z_{\q})}}\frac{m_{\q}}{E_{\q}},  
\label{2}
\eeq   
where $z_{\q}$ is  the  energy fraction
carried by the gluon ($E_{\gl}=z_{\q} E_{\q}$).  
For  light quarks, the opening angle is controlled
by minimum (maximum) values of $z_q$, which in turn
are  determined by the QCD cut-off $Q_0$ 
($Q_0=1$ GeV for JETSET default).  
For heavy quarks, this cut-off
is less important due to the massive factor in (\ref{2}).

Fig.~\ref{see1} illustrates  the angular distribution
of  the  first gluon emission  
with respect to  the BGF quark (antiquark).   
The MC samples 
contain at least one BGF quark  at  $P_T>1.5$ GeV in the 
laboratory frame.  The angle $\Theta_{\q}$   
is  determined as 
$\Theta_{\q}=\arctan ( \vec{p_{\q}}\vec{p_{\gl}}/|p_{\q}||p_{\gl}|)$,  
with $\vec{p}_{\q}$ and $\vec{p}_{\gl}$ being the 
3-momenta of the BGF quark (antiquark)  
and of the parton-shower gluon originating from it. 
The difference between the angular distributions of $\cq$ and $\uq$
quarks illustrates the dead cone effect.
As a cross-check\footnote{The study of the dead
cone phenomenon is complicated by the fact that the most popular
MC models, such as PYTHIA/JETSET, ARIADNE and  HERWIG, do not contain
a switch to exclude the dead cone from their parton showers.}, 
the angular distributions were obtained with
a smaller mass of the charm quark (not shown). In this case
the distribution shown with closed symbols exhibits  at low
$\Theta_{\mathrm{q}}$ much the same rise    
as that for the light quark (open symbols).      
This trend reflects a reduction of the dead-cone size.

To demonstrate  the sensitivity of 
this distribution to the cut-off value, 
the same  distributions are shown for a smaller $Q_0$. The
most important observation is that  
the cut-off does not play significant role
for the gluon bremsstrahlung off the charm quarks,  
while a small  ''dead cone`` in case of the light quark is due to   
kinematic constraints controlled by the value of  $Q_0$.    

\begin{figure}
\begin{center}
\mbox{\epsfig{file=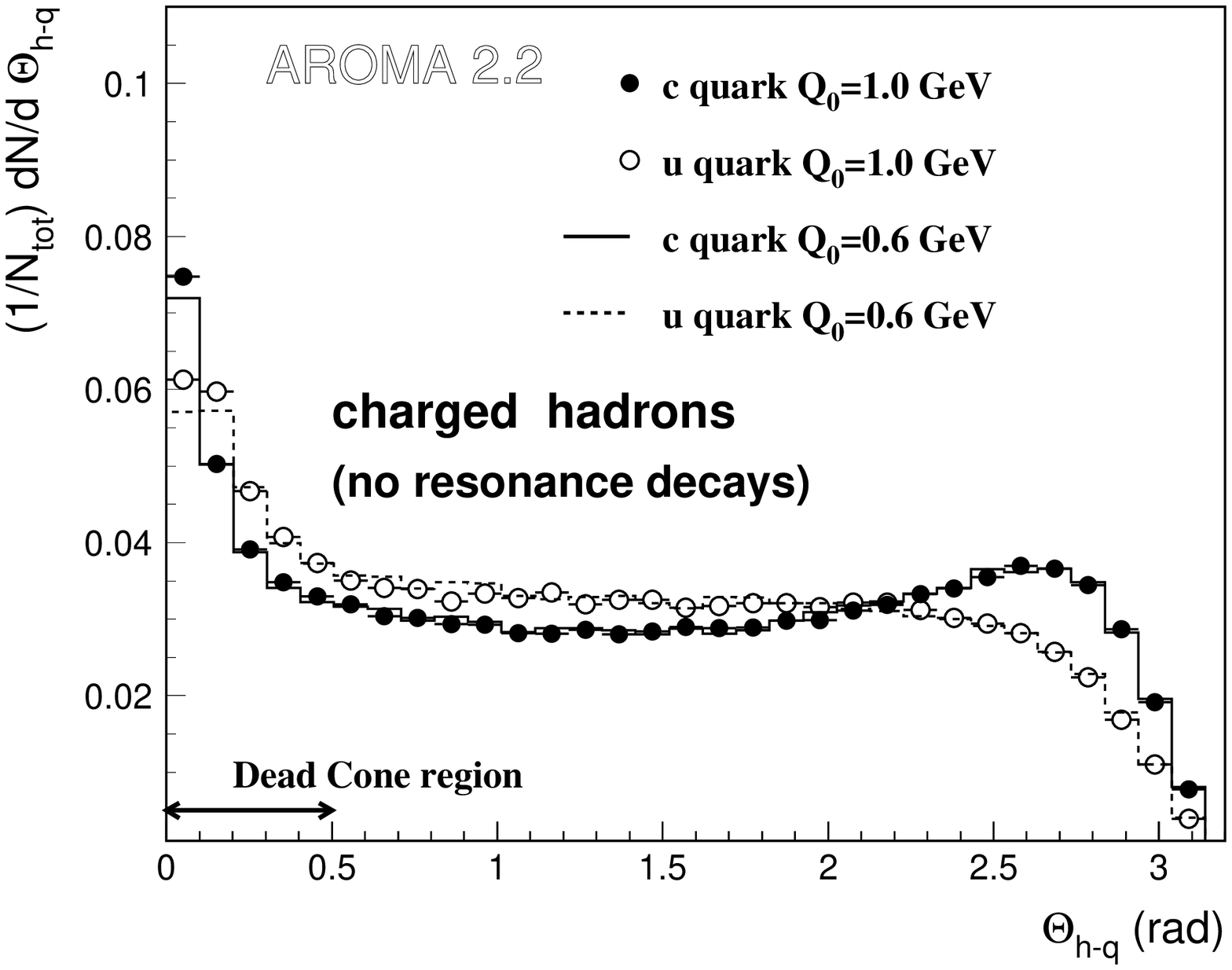,height=10.0cm}}
\caption{\it
Distributions of the angle between the charged hadrons
(without resonance decays) and   
$\cq$ ($\uq$)  BGF quark  for different
cut-off values in the  parton shower.} 
\label{see7}
\mbox{\epsfig{file=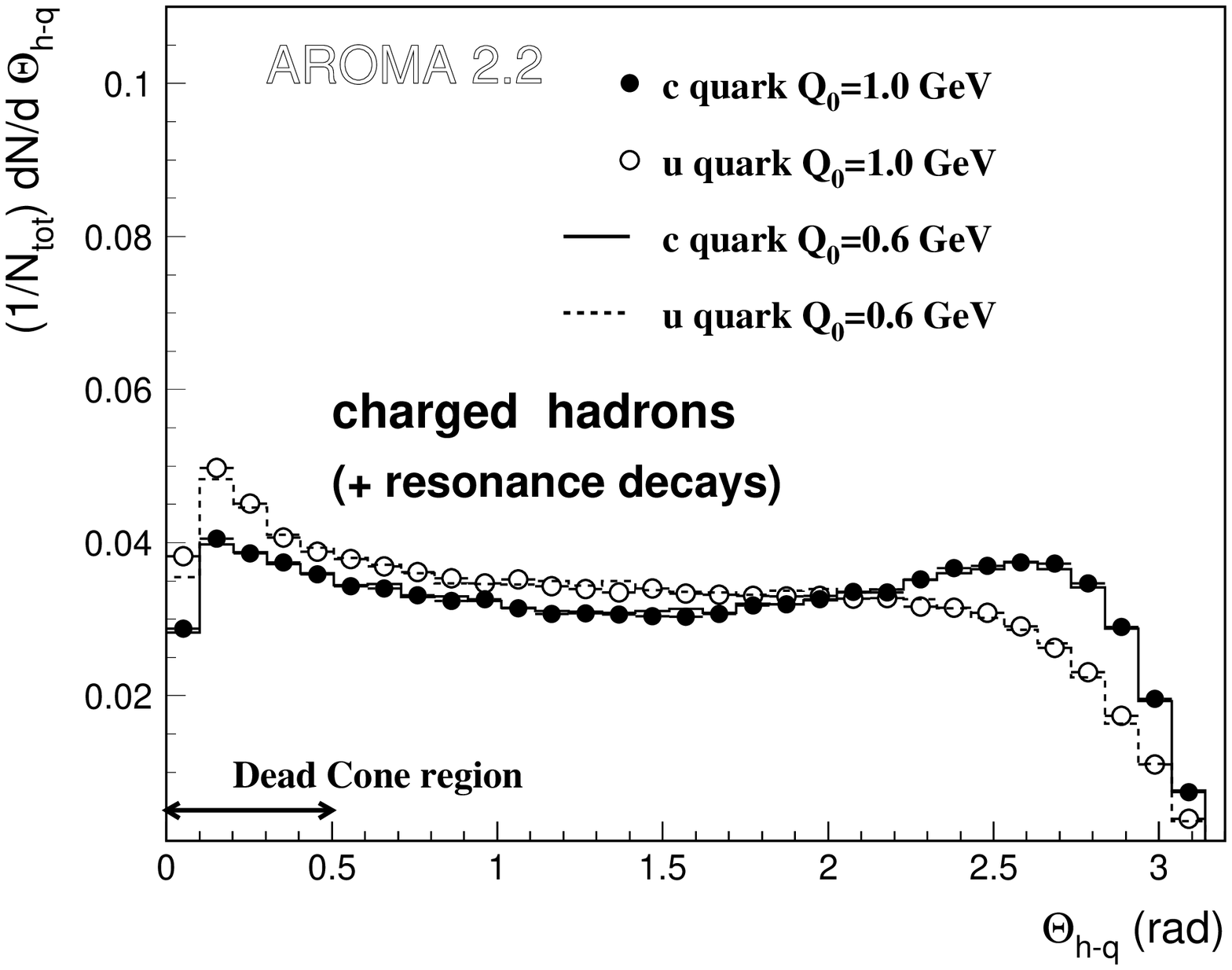,height=10.0cm}}
\caption{\it
Distributions of the angle between the final-state charged hadrons
and $\cq$ ($\uq$)  quark in the BGF for different
cut-off values in the parton shower.}
\label{see3}
\end{center}
\end{figure}

The particle flows around the BGF quarks are shown in
Figs.~\ref{see2}, \ref{see7} and \ref{see3}. The first figure
illustrates   the distributions of the angle between  all  parton-shower
gluons and the original quark.
As for the single-gluon emission shown in Fig.~\ref{see1},  
the dead cone is clearly seen  at $\Theta <0.6$ rad, although
there exists a smearing effect caused by multi-gluon branchings.

The interpretation of Figs.~\ref{see7} and \ref{see3} is more
complicated due to the LUND  string hadronisation involved and
because only charged hadrons are counted.
For the studies reported here only final-state 
hadrons within  $\mid \eta \mid <3$ are used  to avoid counting  
particles close to the beam direction.
Fig.~\ref{see7} shows the angular distribution of charged
hadrons without resonance decays. The dead cone
is not seen anymore.  
One obvious reason for this
is a double counting effect, i.e. a  leading hadron
associated  with  the initial BGF quark contributes to the  
distribution at small  angular separations. 
For charm events, the leading particle is
a charmed  meson which closely  follows to the direction of the
initial $\cq$ ($\cqb$) quark and thus 
contributes to the angular distribution
at  $\Theta_{h-\q}\sim 0$ as seen from  Fig.~ \ref{see7}. 
The resonance production smears  the leading-particle effect 
as shown in  Fig.~\ref{see3}. Another observation
is that the hadron-level  distributions are rather
insensitive  to the QCD cut-off.

\begin{figure}
\begin{center}
\mbox{\epsfig{file=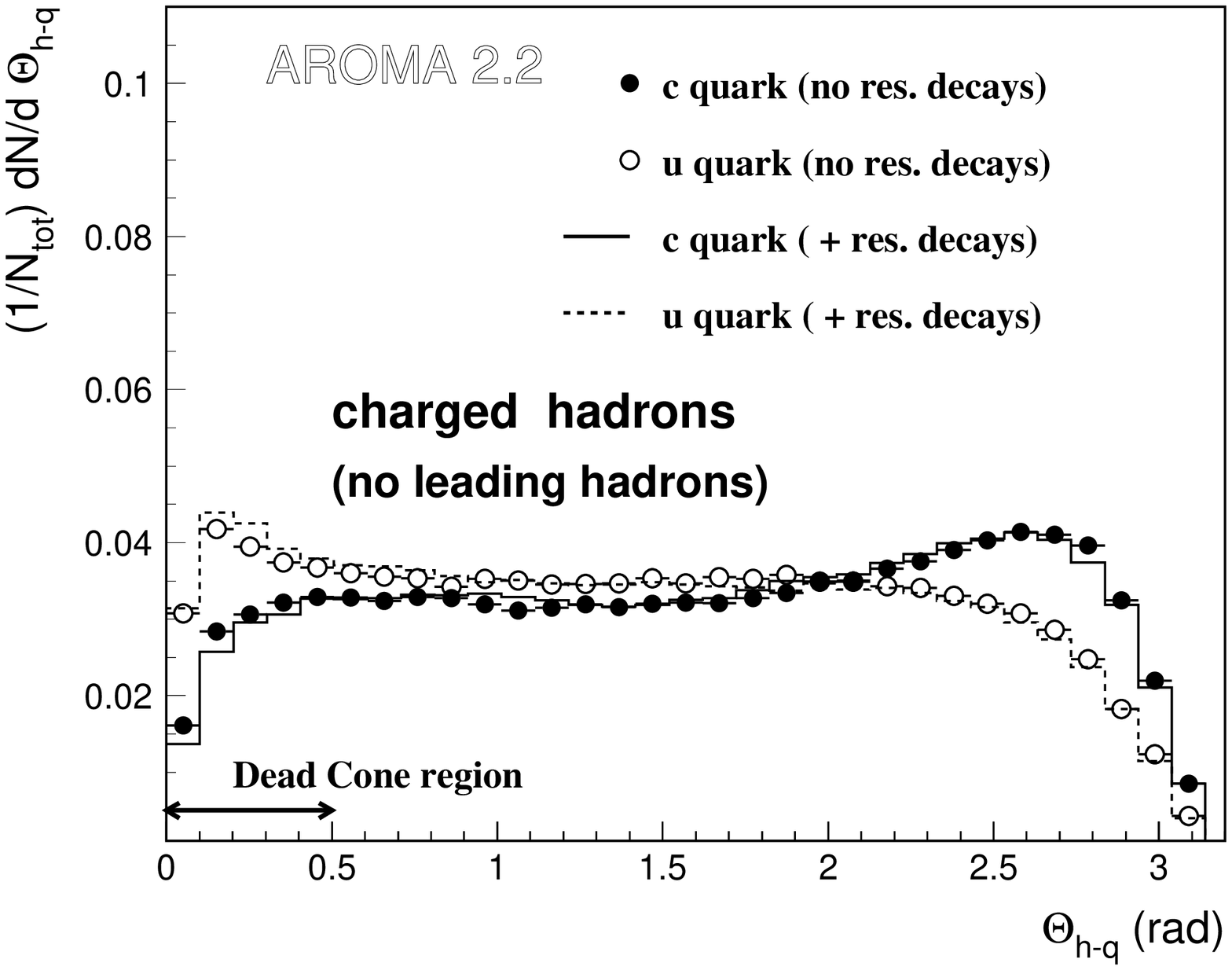,height=10.0cm}}
\caption{\it
Angular distributions for the final-state charged hadrons
(without  leading hadrons)  with respect to   
$\cq$ ($\uq$)  quark in the BGF processes. We show the distributions 
before and after resonance decays 
(symbols and lines, respectively).}  
\label{see8}
\end{center}
\end{figure}

Generally speaking, 
it is not meaningful to say that a given  hadron   
originates from a particular quark in the string fragmentation model. 
However, to investigate the dead cone in more details, 
one can remove the leading
particles containing flavours of the original BGF quark.      
Note that this  can be done in a MC simulation, while the meaning
of the leading particle is more problematic when one
deals with the real experimental situations \cite{chek}.    
Fig.~\ref{see8} shows the angular distribution of the charged
hadrons after removing the leading particles. The figure   
demonstrates  that, despite the smearing effects
from  the parton shower and hadronisation,  the dead cone effect
can be  seen. Moreover, the hadron-level distributions 
for $\cq\cqb$ are  rather close to the ones for parton-level.   

\section{Initial-quark direction reconstruction}

From the previous section it is clear that the most
suitable Monte-Carlo independent 
method to detect the dead cone is to measure  the 
angular distributions of final-state
particles with respect to the original quark. 
To do this, the first step would be to understand  how  well
the initial-quark direction can be reconstructed from
the final-state hadrons.
 
The reconstruction  of  the light-quark direction
can be performed  using a jet clustering algorithm.
The purity for light-quark initiated 
jets in an inclusive DIS event sample can be low, thus
some cuts to reject heavy-flavour events can be used. 
For example, a selection of single-jet DIS events or the 
use of specific cuts \cite{chek} in the Breit frame  to
reduce the BGF type of events  can be useful to suppress 
the contribution of heavy quarks.   

One can measure  the 
direction of the  charm  quark using a few methods. 
One can reconstruct, for instance,  the four-momenta of 
$D^*$ mesons, or to  perform
the clustering of the final-state particles into jets similar
to the light-quark sample. 

We generated two
separate DIS  samples to study the reconstruction  
of the initial-quark direction.  
The first sample contains only $\uq\uqb$  quark topology, while 
the second sample consists of $\cq\cqb$  events from the BGF. 
We require to have at least one quark with a transverse
momentum  larger than 1.5 GeV. In addition, we accept final-state charged 
hadrons within  $\mid \eta \mid <3$.

To reconstruct jets, we use the inclusive 
KTCLUS jet algorithm \cite{ktcl}  with
$P_T$ recombination scheme in the laboratory frame. 
This algorithm is expected to be  
least affected by hadronisation effects and  proton remnants.
We select only a single jet with  highest $E_T$. 

\begin{figure}
\begin{center}
\mbox{\epsfig{file=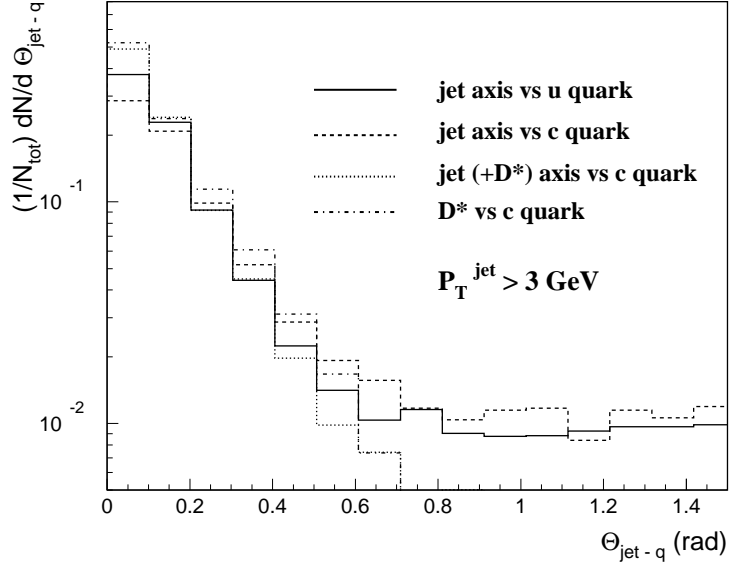,height=10.0cm}}
\caption{\it
Distributions  of the  angle between  the initial $\cq$ ($\uq$) quark
and the jet axis reconstructed using the KTCLUS algorithm for AROMA MC.
The doted line shows the distribution
for events with stable $D^*$ mesons.  
Also shown is the angular distribution  between
the $D^*$ and charm quark (dot-dashed lines).}
\label{see6}
\end{center}
\end{figure}

The solid line in Fig.~\ref{see6} 
shows the angular distribution for the  $\uq\uqb$ sample.   
It is seen that the jet axis   
well reproduces the direction of the original quark. 
Other jet algorithms were  found to be less
reliable than the KTCLUS.

For the $\cq\cqb$  sample, we performed an  analogous study  
(dashed line).   
In this case the jet algorithm cannot reproduce  
the direction of the initial quark as good as for the light quarks.  
The reason for this is decay products of charmed mesons   
as well as  
broadness of the heavy-flavour jets caused by the angular screening. 
Both effects are expected to 
lead to misreconstructions  of the initial-quark direction when 
charged particle multiplicities are small.  

To improve the initial-quark direction  reconstruction 
for the $\cq\cqb$ sample,
we selected a subsample with $D^*$ mesons  and inhibited
their  decays (doted line in Fig.~\ref{see6}). 
The reconstruction of the quark direction 
for this method is as good as for the approach when   
the $D^*$ meson is used to determine the quark direction 
without the jet algorithm (dot-dashed lines).

This  study illustrates that, for the given Monte Carlo
simulation,  the jet axis 
gives a measurement of the original-quark direction to  within
130 mrad ($7^o$) for  the light-quark sample
and  to  within 100 mrad ($6^o$) for the charm sample with stable $D^*$ meson.
The reconstructed $D^*$ meson  itself gives  
the direction of the original  $\cq$  quark to within   80 mrad ($5^o$).  
These values are small,   
compared to the size of the dead cone expected at HERA. Thus,   
the jet axis can be used  to study  the particle flows close to   
light or  heavy quarks.  

\section{Dead-Cone reconstruction}

From the above illustrations it can be seen  that the experimental
observation
of the dead cone is possible by comparing  
the  angular distributions 
of the final-state hadrons around the
jet axis. The study of particle flows close to $D^*$ is the  
most reliable, however,  it is difficult to find a proper
way to compare this measurement with the  gluon 
bremsstrahlung off a light-quark.  
Therefore, the jet reconstruction is required 
both for light- and heavy-quark samples.

\begin{figure}
\begin{center}
\mbox{\epsfig{file=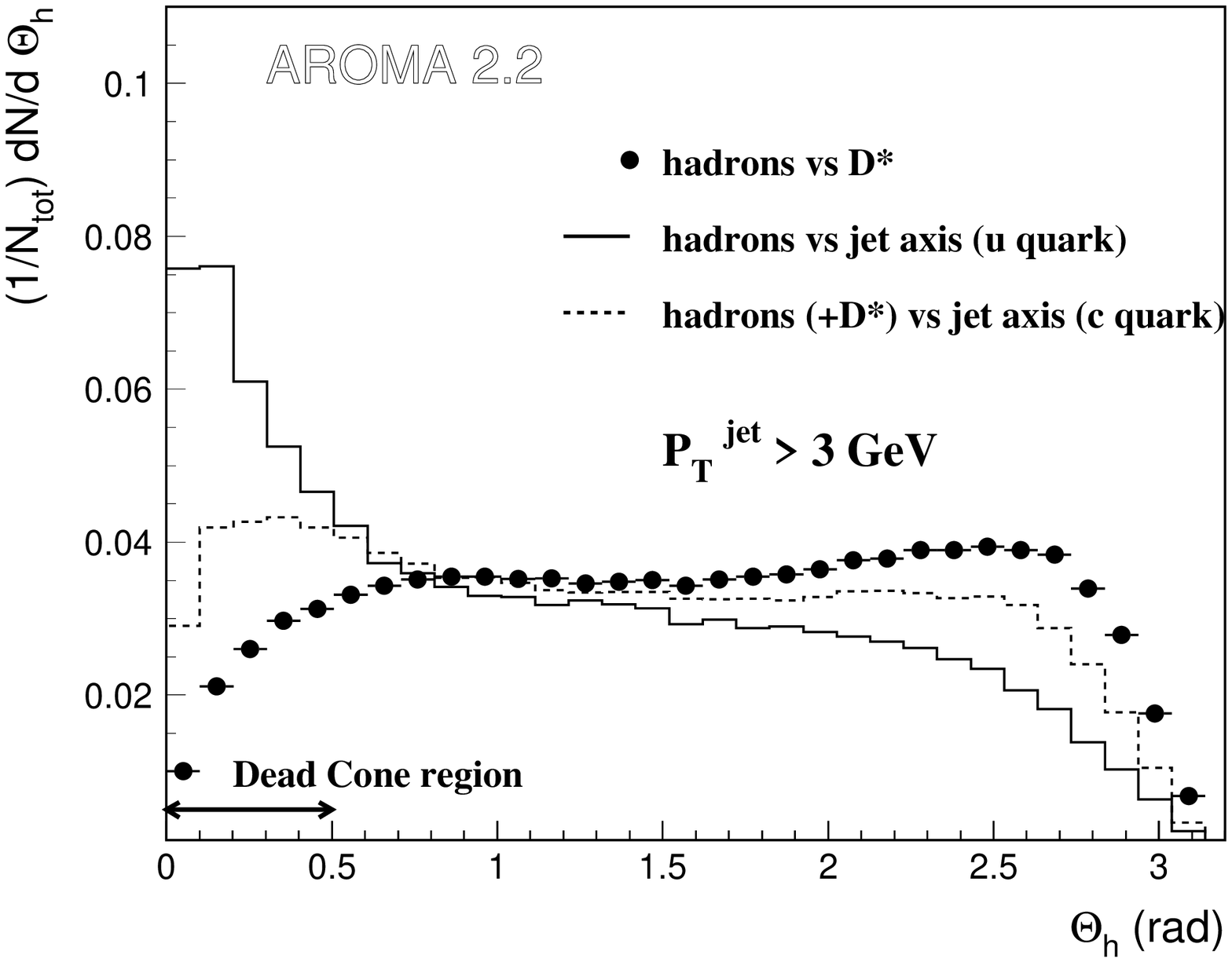,height=10.0cm}}
\caption{\it
Angular separations  between  the final-state hadrons and the jet axis. 
The solid line shows the distribution 
of hadrons with respect to the jet axis for a light quark sample,  
while the dashed line shows the  distribution for the $\cq$-quark sample
with stable $D^*$ mesons.  Also shown is 
the angular  distribution of the charged particles
with respect to the direction of motion of the $D^*$ meson.}
\label{see5}
\end{center}
\end{figure}

\begin{figure}
\begin{center}
\mbox{\epsfig{file=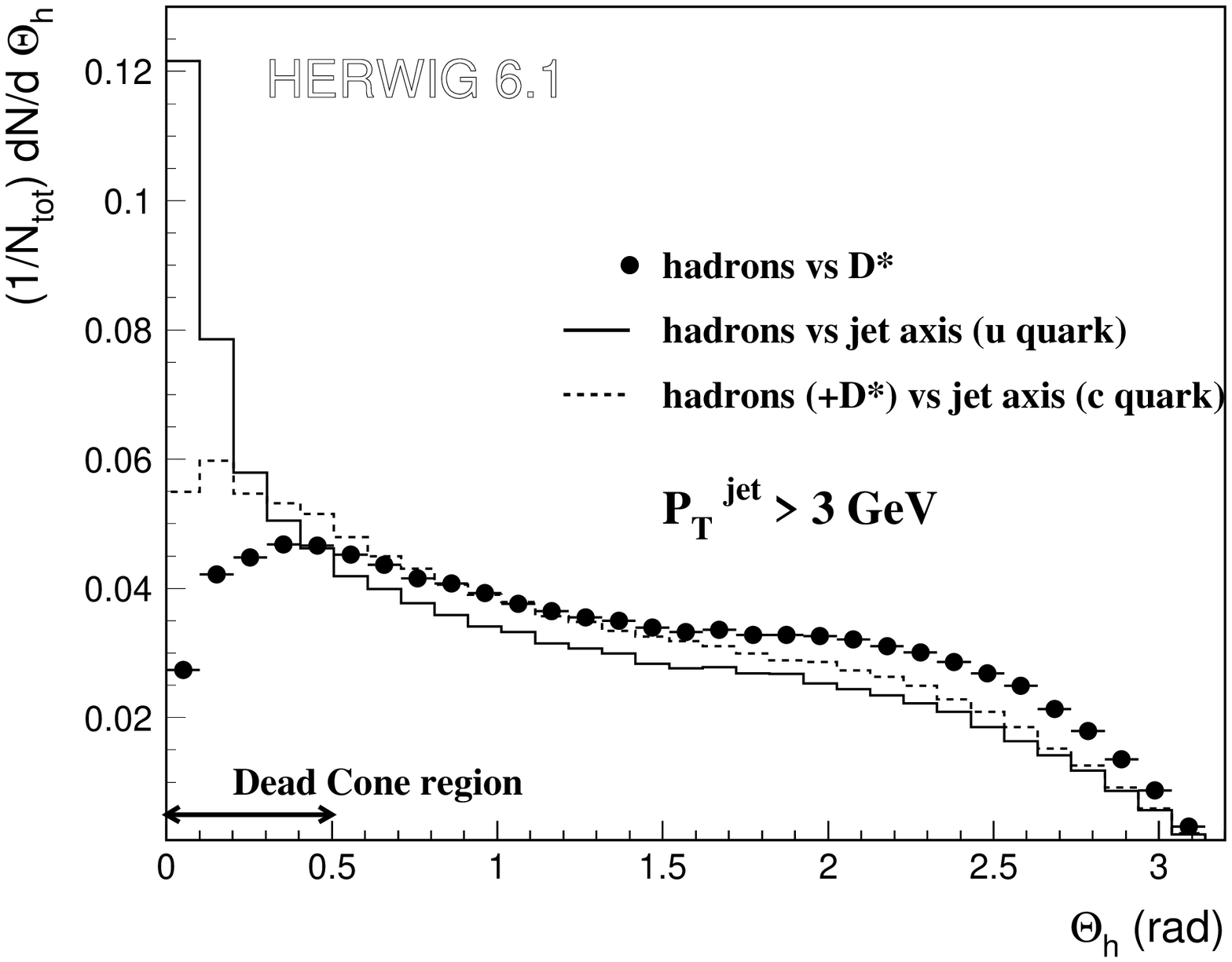,height=10.0cm}}
\caption{\it
Same as Fig.~\ref{see5}, but for the HERWIG 6.1 model.}  
\label{see5h}
\end{center}
\end{figure}

\vspace{0.3cm}
The dead cone  can be measured in a few steps:

\vspace{0.3cm}
1) To select  inclusive events
for a given phase-space region in $Q^2$ and $x$ and 
to reconstruct the jet axis using a  jet algorithm,
preferably the KTCLUS. 
Then the angular distribution
of charged tracks with respect to  the 
jet axis should be found.
Note that additional  cuts \cite{chek} to reject the BGF type of events
can be useful to reduce contributions from  heavy quarks.   

\vspace{0.3cm}
2) To select a sample of events with reconstructed 
$D^*$ mesons for the same kinematic region
as for the inclusive sample enriched with light quarks. 
To identify the charmed mesons, one can use the most
popular exclusive decay channel   
$D^{*\pm} (2010)\to D^0\pi^{\pm}_s$ with
$D^0\to K^-\pi^+$ ($+$c.c.), where $\pi^{\pm}_s$ refers to a low
momentum ("slow")  pion.
The decay products of the $D^*$ should be rejected and 
4-momenta  of the reconstructed $D^*$ mesons should be added 
to the same events.

\vspace{0.3cm}
3) To apply the jet cluster algorithm to
the obtained sample using the same 
$P_T$ cut as for the inclusive events.  
The reconstructed jet axis can be used to obtain
the angular distribution of all final-state hadrons  
with respect to the initial BGF quark.   

\vspace{0.3cm}
4) To compare the angular distributions  for the light-quark
and  charm samples. A  difference between these two 
distributions at low $\Theta$ should 
give an estimate of the dead cone.  
  
\vspace{0.3cm}
Fig.~\ref{see5} shows the angular distributions using
the prescription described above. We use $P_T >3$ GeV cut
to find jets both for 
the light-flavour and $\cq\cqb$ event samples. The latter contains  
the $4$-momenta of stable  $D^*$ mesons.  
The difference between the solid and the dashed lines  gives an  
estimate of the dead-cone size ($\sim 25-30^o$). 
Note that this difference is
bigger than that for the parton-level distributions due to 
the jet algorithm used to determine the initial-quark direction.

It should be noted that the $D$-meson fragmentation function
in JETSET is a function of the charm mass and is harder
than that for light-quark mesons. Therefore, the dead cone effect
can be influenced by  non-perturbative effects  related to the
fragmentation \cite{Sro}.   
As a final check on the above result, Fig.~\ref{see5h} shows
the same distributions  but using the HERWIG 6.1 \cite{HRW} model which is
based on the cluster hadronisation scheme. The model was
used with default parameters and   
the DIS events were generated for the same kinematic
range as for the AROMA. The  HERWIG angular 
distributions are shifted to low $\Theta_{\mathrm{h}}$ values, but 
are found to be a similar shape to those shown in Fig.~\ref{see5}.
The difference between the solid and 
dashed lines seen from  Fig.~\ref{see5h} again 
gives an estimate of the dead-cone size,  
which is smaller than that of the AROMA.
This result may indicate that there exists a contribution to the observed
depletion from fragmentation mechanism, however, in this
paper we refrain from attempting a more detailed study of this issue.

In conclusion, a Monte-Carlo independent 
method to observe the dead cone for
DIS or photoproduction is proposed. 
It is based on the measurement of particle flows close 
to  the jet axis for heavy and light-quark samples.   
Using  Monte Carlo simulations, it is shown that such a 
study worths experimental investigations,  providing   
an estimate of the dead-cone size which is of 
importance for discriminating  different parton-shower
implementations, especially with respect to  inclusion  of 
heavy quarks.

\section*{Acknowledgments}
I am indebted to W.Ochs for stimulating discussions
on this topic.  
I also thank M.~Derrick, E.~De.~Wolf, L.~Gladilin, L.~L\"{o}nnblad,  
J.~Repond and T.~Sj\"ostrand for useful comments. 

{}

\end{document}